\documentclass[aps,pre,twocolumn,showpacs,floatfix,preprintnumbers,amsmath,amssymb]{revtex4}
\usepackage{epsfig}
\usepackage{graphicx}
\usepackage{dcolumn}
\usepackage{bm}

\begin{document}
\title{Manipulation of extreme events on scale-free networks}
\author{Vimal Kishore, Abhijeet R. Sonawane\footnote{Present address : FIRST, Collaborative Research Center 
for Innovative Mathematical Modelling, Institute of Industrial Science, The University of Tokyo.}
and M. S. Santhanam}
\affiliation{Indian Institute of Science Education and Research,\\
Homi Bhabha Road, Pune 411 021, India.}
\date{\today}

\begin{abstract}
Extreme events taking place on networks are not uncommon.
We show that it is possible to manipulate the extreme events occurrence probabilities
and its distribution over the nodes on scale-free networks by tuning the
nodal capacity. This can be used to reduce the number of extreme events
occurrences on a network. However, monotonic nodal capacity enhancements, beyond a point,
does not lead to any substantial reduction in the number of extreme events. We point out
the practical implication of this result for network design in the context of
reducing extreme events occurrences.

\end{abstract}
\pacs{05.45.-a, 03.67.Mn, 05.45.Mt}
\maketitle

\section{Introduction}
The study of dynamical processes \cite{vesp} and extreme events \cite{netxv,netxv1} on complex networks
has become an important topic of research interest both for its inherent scientific
understanding and possible applications. The main motivation being that many
extreme events such as the
floods, congestion in internet and other communication networks, power black-outs and traffic
jams take place on complex networks. In particular, the effect of these
extreme events is tremendously high in terms of its impact on human lives, property and
productivity \cite{eevent1}.
This is evident, for instance, in the life and property lost due to floods and
the working hours lost due to power black-outs and traffic jams \cite{prod}.
It would be beneficial if some form of control or at least a possibility of manipulating
the occurrence of extreme events can be achieved. In this work, we show
that we can manipulate the extreme event occurrence probabilities
on networks and we examine the extent to which this is feasible.

We assume that a suitable model for transport is defined on any network such that
the events in any of its node is proportional to the flux $x(t)$ passing
through it.
For instance, flux could be the volume of water flowing in a
river or the number of information packets passing through a router in an IP network.
Extreme events are associated with exceedences of the flux above a certain threshold, {\it i.e.}, 
an event is called an extreme event at time $T$ if
$x(t=T) \ge x_{ee}$ where $x_{ee}$ is the threshold used to identify extreme events.
In this setting, we focus on the extremes, arising due to
inherent fluctuations, in the flux. Clearly, then the
extreme events cannot be avoided altogether in any finite system. They can 
at best be partially mitigated by tuning a suitable network parameter.
This is the theme we explore in this work. As this approach allows some freedom to
manipulate EE on networks, it is similar in spirit to the
emerging interest in controlling the dynamics on complex networks \cite{control}.

Our transport model is the dynamics of random walkers on networks.
Though random walk on networks were studied earlier \cite{rw1,rw2}, they
were not focussed on the extreme event (EE) properties. In a recent work \cite{netxv,netxv1}, the
threshold for extreme events $x_{ee}$ was taken to be proportional to the typical
size of the flux passing through a node, i.e,
$x_{ee} \propto \sigma_f = \langle f \rangle^{1/2}$, where $\langle f \rangle$ and $\sigma_f$
are the mean and the standard deviation of the flux passing through the node.
In the present work, we choose the threshold to be
\begin{equation}
x_{ee} \propto \sigma_f = \langle f \rangle^{\alpha}, \;\;\;\; \alpha \ge 0.
\end{equation}
Main motivation for this choice arises from the
results in Ref. \cite{ym} which show that $\sigma_f \propto \langle f \rangle^{\alpha}$ 
and in particular the value of $\alpha$ depends on factors such as the resolution
of the flux measurements and the noise in the number of walkers.

Another motivation arises because $x_{ee}$ can be interpreted as the nodal capacity \cite{barrat}
and extreme events occur when flux exceeds the nodal capacity. The empirical and
modeling studies \cite{motter1} on the relationship between {\sl mean} load and {\sl mean} capacity
in a network
do not address the question of extremes in load fluctuations. Extreme events
result from large fluctuations in load and these are mostly responsible for temporary
local network failures.
Our experience suggests that most of such extreme events, say the vehicles
piling up at a traffic intersection, arise due to limited throughput capacity
of the nodes. Then, one possible solution to reducing
the extreme event occurrences would be to increase the handling capacity of the node.
Guided by this intuition, we vary $\alpha$ which is a proxy for the nodal capacity such
that larger values of $\alpha$ correspond to larger handling capacity of the node.
In this work, we show that it is possible
to manipulate the probabilities for the occurrence of extreme events on the
nodes of a scale-free network by tuning $\alpha$.

A more significant result is that tuning $\alpha$ beyond a certain point
does not lead to any significant reduction in the number
of extreme events.  This result has important implications
when we opt for capacity building as the route to mitigate extreme events on networks. Capacity
addition invariably comes at a heavy price. For instance, building bridges at a road
intersection is one method of capacity addition that might increase the throughput
across the node. In communication networks, additional servers and switches might be
needed at a node to smoothly handle excess traffic flowing through it. All these
interventions require expensive infrastructural changes. In view of such high costs
involved in this effort, 
it is important to ask if increase in capacity will lead to proportionate decrease
in the likelihood of extreme events. The results in this work furthers our
insight into this question in the context of extreme events on complex networks.

\section{Random walks and extreme events on networks}
We model the flux as a random walk process executed by $W$ independent walkers on a 
fully-connected, unweighted network.
In this case, the distribution of walkers
passing through $i$-th node is given by \cite{netxv,rw1},
\begin{equation}
f_i(w)={W \choose w}~p_i^w~(1-p_i)^{W-w}.
\label{binomial}
\end{equation}
The stationary probability $p_i$ to find a walker on $i$-th node is
\begin{equation}
p_i=\frac{k_i}{\sum_{l=1}^{N}k_l}.
\label{statdist}
\end{equation}
This measures the extent to which the walkers are
attracted to $i$-th node and it depends on its degree.
The flux of walkers being binomially distributed, the mean 
$\langle f_i \rangle$ and variance $\sigma_i^2$
of the flux passing through $i$-th node can be immediately written down 
and both depend only on degree $k_i$.

In the spirit of extreme events statistics \cite{mss}, we define an
event on node $i$ to be extreme if $w > q_i$, the threshold $q_i$ to be
determined below. Then, the probability for the occurrence of
extreme event on node $i$ is given by
\begin{equation}
F_i(q_i) = \sum_{w=\lfloor q_i \rfloor}^W f_i(w) = I_p(\lfloor q_i \rfloor + 1, W - \lfloor q_i \rfloor ),
\label{xvp}
\end{equation}
where $\lfloor . \rfloor$ is the floor function and $I_z(a,b)$ the incomplete Beta function \cite{abs}.
We note that the extreme event probability depends on the choice of threshold $q_i$
and this serves as a handle to tune and control extreme event probability $\mathcal{F}_i$.
In this work, the threshold is chosen as
\begin{equation}
q_i = \langle f_i \rangle + m \sigma_i^\alpha \approx
W \frac{k_i}{\sum_{l=1}^{N} k_l} + m\left( W \frac{k_i}{\sum_{l=1}^{N} k_l} \right)^{\alpha},
\label{thresh2}
\end{equation}
where $m, \alpha \ge 0$ are real numbers. The magnitude of extreme event $m$
scales the probability for extreme events \cite{netxv} and does not qualitatively change it.
Hence, $\alpha$ provides a handle to manipulate the extreme events by tuning the threshold $q_i$.

\subsection{Manipulation of extreme events}
\begin{figure}
\centerline{\includegraphics*[width=3in,angle=0]{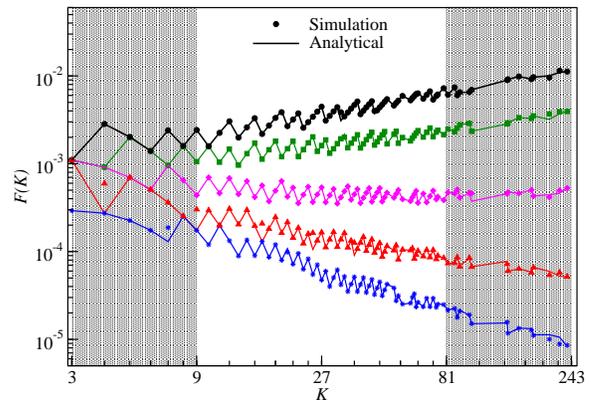}}
\caption{The probability for the occurrence of extreme events as a function
of degree $k$, shown as log-log plot, for $\alpha=0.40, 0.43, 0.47, 0.50, 0.52$. Both
the analytical (Eq. \ref{xvp}) and simulation results are shown. The shaded region on the left (right)
corresponds to small degree nodes (hubs).}
\label{fig1}
\end{figure}

To illustrate the idea of manipulating the extreme events, we display the probability
for the occurrence of extreme events in Fig. \ref{fig1}. The simulation
results shown in this figure (and in the rest of the paper) are obtained from
random walk simulations on a scale-free network (degree exponent : $\gamma = 2.2$)
of size $N=5000$ and $W=39830$ averaged over 100 realizations for each value of $\alpha$.
Clearly, there is a good agreement between the analytical (Eq. \ref{xvp}) and simulation results.
Note that as $\alpha$ is increased from 0.4 to 0.6, the probability for the occurrence
of extreme events $F_i(k)$ changes systematically. For the hubs ($k > 80$), $F_i$ changes
by nearly 2-3 orders in magnitude. A significant feature is that for $\alpha < 0.47$, extreme event
probability $F_i$ is higher, on an average, for the hubs $(k > 80)$ when
compared to the small degree nodes $(k < 20)$. This feature reverses for $\alpha > 0.47$,
i.e., $F_i$ is higher for small degree nodes compared to that of hubs.
For $\alpha \approx 0.47$, the probability
$F_i$ is approximately independent of the degree of the node (ignoring the local fluctuations).
By tuning $\alpha$ we obtain a range of behaviour for the
EE occurrence probabilities.

Physically, results in Fig \ref{fig1} can be interpreted as follows. Note that 
$\alpha=1/2$ and $m=1$ in Eq. \ref{thresh2} represents the typical size $w_{typ}$ of flux
through a node in the random walk environment. We can take the threshold
$q_i$ to represent the capacity of the node to handle this flux $w_{typ}$.
Then, $\alpha < 1/2$ would imply $w_{typ} > q_i$, i.e, flux is more than
the capacity of the node to handle it. This can be thought of
as the congestion-like situation.
On the other hand, $\alpha > 1/2$ corresponds
to $w_{typ} < q_i$, in which typical size of flux is
smaller than the capacity of the node and hence congestion-like situation is less likely
to happen.
Fig. \ref{fig1} implies that as $\alpha$ is varied, the probability of EE
on the hubs shows larger variability than the small degree nodes (compare the shaded regions
on the left and the right). Hence, we have far less control over the EE taking
place on the small degree nodes. This is not the case for the hubs. The total number of EE on
the entire network will depend on the events taking place on both the hubs and the small degree nodes.
Thus, by adjusting the capacity or tuning $\alpha$, we can manipulate the
EE taking place on scale-free network. 

\begin{figure}
\centerline{\includegraphics*[width=2.7in,angle=0]{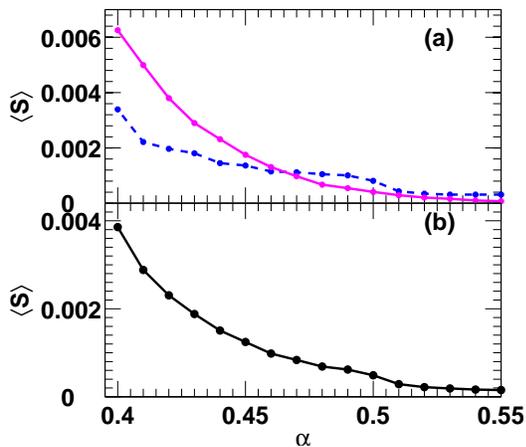}}
\caption{(a) The mean queue size $\langle S \rangle$ for the small degree nodes (dashed line)
and the hubs (solid line),
(b) the mean queue size for the entire network $\langle S \rangle$ plotted
as a function of $\alpha$. Note that larger $\alpha$ correspond
to larger nodal capacity.}
\label{fig2}
\end{figure}

\subsection{Excess load as queue size}
In any real network that encounters congestion-like situation, the traffic in excess of the 
nodal capacity leads to a build-up of queue. For instance,
clustering of vehicles at the traffic junctions or pending http requests to a web server or
phone connections waiting to be serviced by cellular hubs are all
examples of such a build-up. If $\alpha < 1/2$, we should expect to see such queue and the number
of walkers waiting in that hypothetical queue or buffer [] is an indicator of the severity of the extreme event.
In models in which queue does not get cleared, under certain conditions, jamming or
a cessation of dynamics can take place. Phase transition to such a jamming state has been
well studied \cite{jam}.
If $w_i(t)$ represents the number of walkers on $i$-th node at times $t=1,2,...T$,
then the queue length is $Q_i(t)= (w_i(t)-q_i) \theta\left(w_i(t)-q_i\right)$. In this,
unit step function $\theta(.)$ is used to ensure that $Q_i(t)=0$ whenever $w_i(t) < q_i$.
We define the mean queue size $\langle S \rangle$ for a network with $N$ nodes as
\begin{equation}
\langle S \rangle = \lim_{T\to\infty} \frac{1}{TN} \sum_{i=1}^N \sum_{t=1}^T Q_i(t)
                  =  \frac{1}{N} \sum_{i=1}^N f_i(w) Q_i
\end{equation}
In this, $\langle S \rangle$ measures the mean number of walkers, {\it in excess of nodal capacity},
present in the queue per node at every time instant.
Larger values of $\langle S \rangle$ imply more congestion-like scenario in the network.
Now, we can obtain a broader picture of control over the extreme events as a
function of $\alpha$. Fig. \ref{fig2}(b) shows the mean queue size $\langle S \rangle$
obtained from random walk simulations with $m=4$ as a function of $\alpha$. 
Evidently, $\langle S \rangle$ decreases with the increase in network capacity.
Note that as argued
before, $\alpha < 0.5$ corresponds to a congestion-like situation and this is evident
from the large values for $\langle S \rangle$. In contrast, for $\alpha > 0.5$,
$\langle S \rangle \approx 0$ indicating almost no
extreme events on the network. Within the frame work of extreme events arising due
to inherent fluctuations in the flux, it is impossible to eliminate them altogether
and hence $\langle S \rangle$ will never be exactly zero.
However, the probability of extreme event occurrence and the value of $\langle S \rangle$
can be made arbitrarily small by tuning $\alpha$.

Even as we manipulate the overall congestion-like scenarios on
a network we explore the finer details.
As Fig. \ref{fig1} reveals, the extreme event probability on the 
hubs and small degree nodes have different
dependence on degree $k$. This leads to an unequal apportioning of the
burden of extreme events on hubs and small degree nodes of a network.
We define a mean degree $\widetilde{k}$ such
that we denote nodes with $k > \widetilde{k}$ as hubs and $k < \widetilde{k}$ 
as small degree nodes.
Fig. \ref{fig2}(a) show $\langle S \rangle$
for hubs (solid line) and small degree nodes (dashed line) separately.
For $\alpha < 0.5$, hubs display larger queue
sizes as compared with the small degree nodes. In a reversal of roles,
for $\alpha > 0.5$, hubs display small queue
sizes when compared with the small degree nodes. Quite interestingly,
at $\alpha \approx 0.47$, both the hubs and the small degree nodes
equally share the burden of extreme events
and queue sizes. This is indicated by the crossing of both
the curves in Fig. \ref{fig2}(b). This is exactly the value of $\alpha$
at which the probability for extreme events becomes independent of degree $k$
in Fig. \ref{fig1}. Irrespective of how the hubs and small degree nodes are
defined, $\alpha \approx 0.47$ would always remain the crossover point.
Thus, except for $\alpha \approx 0.47$, the network does not share the
burden of extreme events in an egalitarian manner among the small degree
nodes and the hubs. The practical implication is that the designing
nodal capacity implicitly affects the 'spatial' distribution of  extreme events
over the nodes of a scale-free network.

\begin{figure}
\centerline{\includegraphics*[width=3in,angle=0]{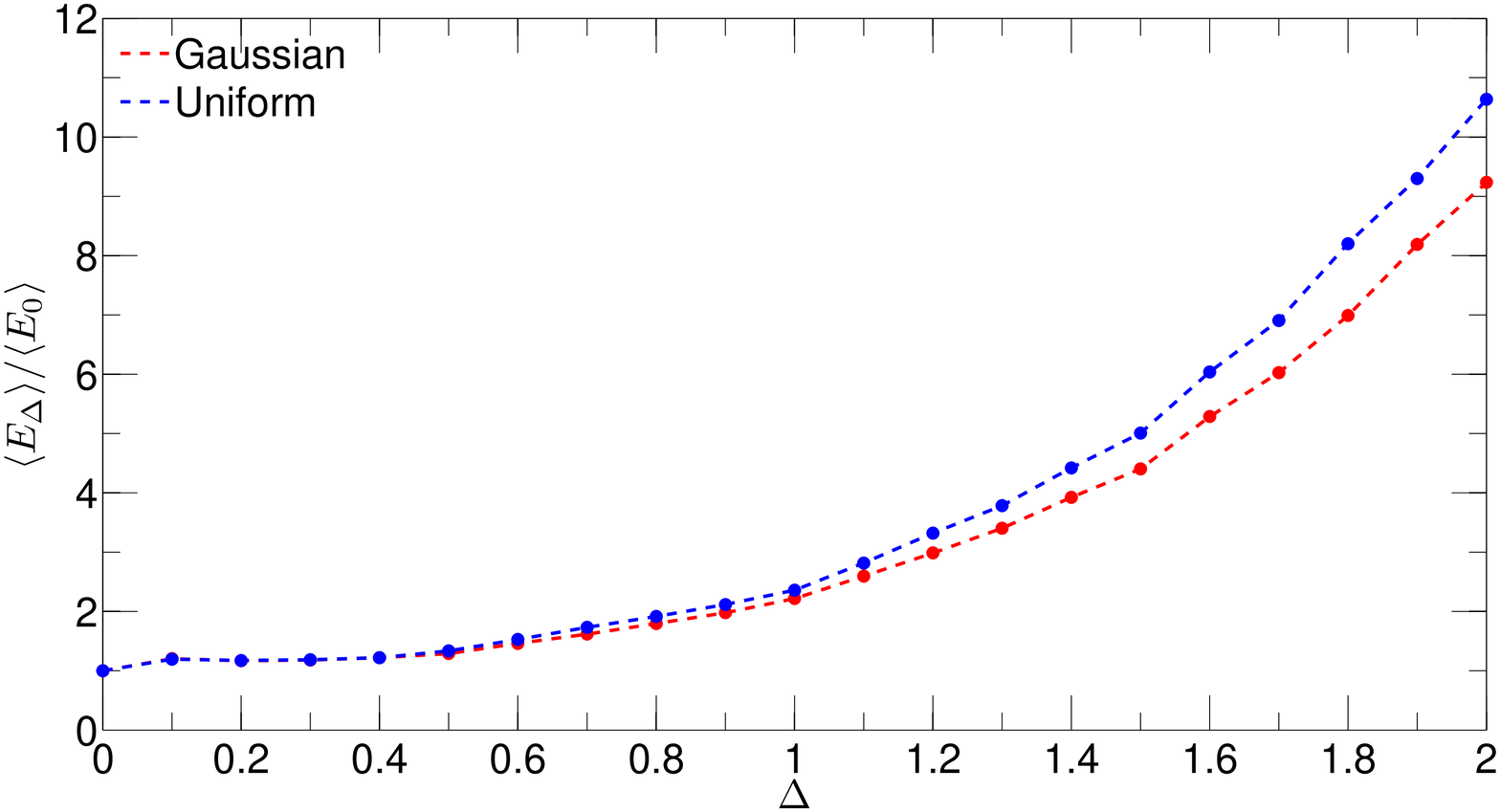}}
\caption{Scaled mean number of extreme events 
$\langle E_{\Delta} \rangle / \langle E_{0} \rangle$ with $m=4$ and $\alpha=1/2$
as a function of strength of noise $\Delta$. Random numbers were
drawn from uniform and Gaussian distribution.}
\label{fig3}
\end{figure}


\section{Effect of variability in nodal capacity}
In this section, we will take the capacity of $i$-th node to be $C_i=q_i$.
The results in Sec. II implicitly assume that the capacity of $i$-th node $C_i$
is related to its degree $k_i$ through Eq. \ref{thresh2} and that all the nodes with identical
degree have identical capacity. However, this condition is almost never satisfied in most 
of the real networks \cite{barrat,motter1}.
In reality, the nodal capacity of $i$-th node does not follow any prescribed formula and indeed
could be treated as a random variable drawn from a suitable probability distribution. Hence, we have
for the time-independent capacity of $i$-th node
\begin{equation}
C_i(\Delta) = \langle f_i \rangle + ( m \sigma_i \pm \Delta \xi_i \sigma_i) = 
C_i(0) \pm \Delta \xi_i \sigma_i
\label{node-cap}
\end{equation}
with $\Delta$ being the strength of noise and $\xi_i$  a random number.
Evidently, if $\Delta=0$, this reduces to Eq. \ref{thresh2}, the threshold for extreme events.
We compute the mean number of extreme events over the entire network, i.e, 
$\langle E_{\Delta} \rangle = (1/NT) \sum_{t=1}^T \sum_{i=1}^{N} \theta(w_i(t)-C_i(\Delta))$,
scaled by $\langle E_{0} \rangle$. Thus, after simple manipulations, we get,
\begin{equation}
\frac{\langle E_{\Delta} \rangle}{\langle E_{0} \rangle} = 
\frac{\sum_{i} F_i(C_i(\Delta))}{\sum_{i} F_i(C_i(0))}
\label{nrat}
\end{equation}
By construction, this quantity is unity at $\Delta=0$.
As simulations in Fig. \ref{fig3} reveal, $\langle E_{\Delta} \rangle / \langle E_{0} \rangle$
for $m=4$ increases nonlinearly as the strength of noise $\Delta$ increases. It agrees
with the analytical result obtained using Eqs. \ref{nrat} and \ref{xvp}.
Notice that Eq. \ref{node-cap} can be written as $C_i(\Delta) = \langle f_i \rangle + \widetilde{m} \sigma_i$ 
where $\widetilde{m} = m \pm \Delta \xi_i$, i.e, every node has a different value of 
capacity indexed by $\widetilde{m}$.
We use the fact that the extreme event probabilities $F_i$
scale with magnitude \cite{netxv} to obtain a leading order estimate of
$\langle E_{\Delta} \rangle/\langle E_{0} \rangle \approx 1 + a_1 \Delta + a_2 \Delta^2$, where
$a_1$ and $a_2$ depends on the degree of the node.
This conclusion is independent of whether
the random numbers $\xi_i$ were drawn from uniform or
Gaussian distribution. 

Physically, this can be understood as follows. Suppose $C_i(0)$ represents $4\sigma$
capacity for all the nodes of the network. For noise strength $\Delta > 0$,
some nodes will have capacity $C_{-}$ such that $C_{-} < C_i(0)$ while others will have
a capacity $C_{+}$  such that $C_{+} > C_i(0)$. The nodes with $C_{-}$ will 
encounter more extreme events
and hence higher probability for the occurrence of extreme events (in comparison with $C_i(0)$ case).
However for nodes with capacity $C_{+}$ the extreme event
occurrence probability is much smaller and leads to fewer events. Hence we notice a net increase
in the number of extreme events on $i$-th node. As $\Delta$ increases, this effect translates into
an increase in the mean number of extreme events over the entire network.
In most of the real-life
networks, the nodal capacity is typically proportional to the flux passing
through the node \cite{betcen} but is unlikely to adhere strictly to specifications such as $C(0)$
assumed in Eq. \ref{node-cap}. Hence it is an appropriate guess that nodal
capacity being proportional to its importance as measured by degree could be its mean
behaviour \cite{barrat} but the actual capacities would be a random variable \cite{cap} clustered around this
mean behaviour. In all such cases, the random walk based model studied here predicts
that the number of extreme events will increase with the variability in the
nodal capacity. For other values of $\alpha$ too qualitatively similar results
as in Fig. \ref{fig3} are obtained.

\begin{figure}
\centerline{\includegraphics*[width=3.5in,angle=0]{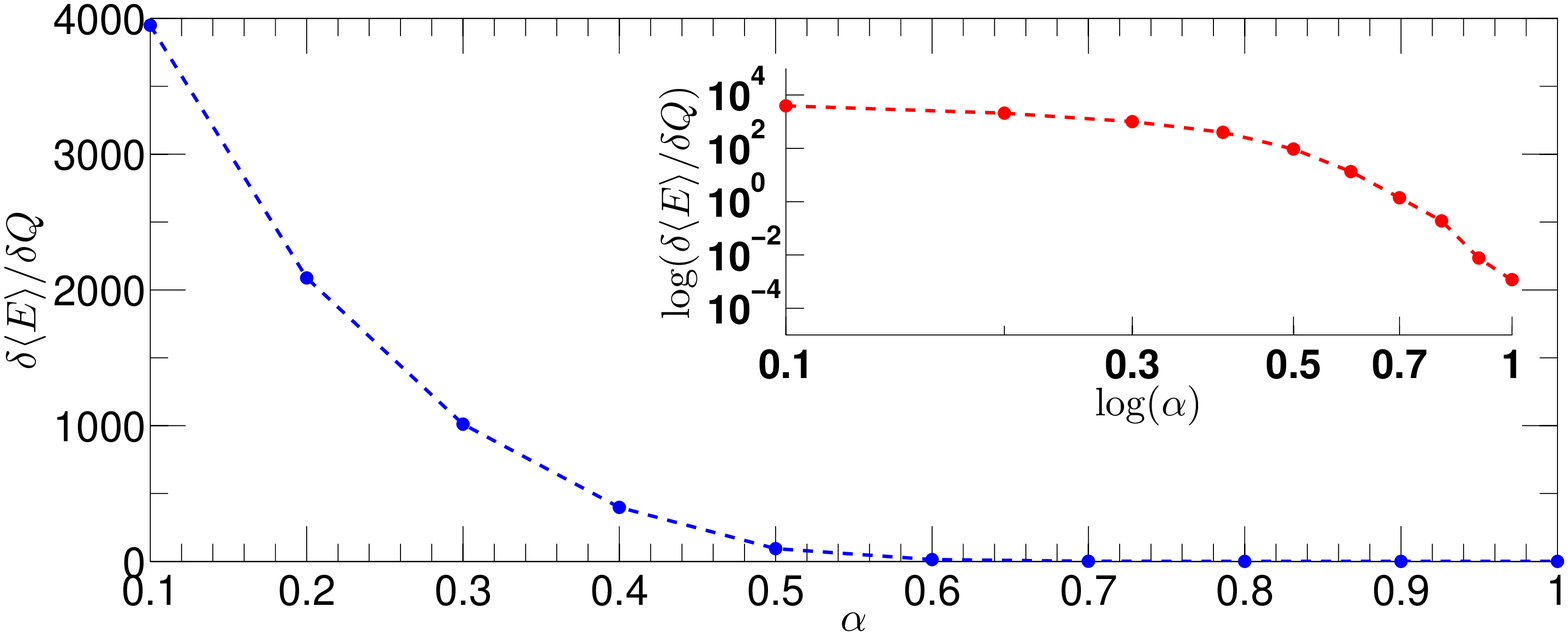}}
\caption{The ratio of change in mean number of extreme events $\delta \langle E \rangle$
on the network for $\delta C$ change in capacity of the network plotted as a function
of $\alpha$. This is obtained from random walk simulations (solid circles) with $m=4$.
The inset shows the same data in log-log plot.}
\label{fig4}
\end{figure}

\section{Capacity addition and extreme events}
Given that the variability in the nodal capacity leads to an overall increase
in the number of extreme events (as shown in Fig. \ref{fig3}), we ask how we can
reduce the mean number
of extreme events on the network. This would be a desirable requirement
for a smooth functioning of the networks \cite{cec}. As argued before, intuitively we expect mean
number of extreme events to decrease if nodal capacity increases. We can increase capacity
by increasing $\alpha$. Further, to simplify the scenario, we take $\Delta=0$ for this part.
Figure \ref{fig4} shows simulation results for the change in
the mean number of extreme
events when capacity changes by one unit, i.e, $\delta \langle E \rangle / \delta C$
as a function of $\alpha$.
For $\alpha < 0.5$, the $\delta \langle E \rangle / \delta C$ curve
decays quickly. In this regime, capacity addition is accompanied
by a large reduction in the number of extreme events.
Hence, any capacity addition in this regime yields rich dividends in terms of cost-benefit ratio.
However,  beyond $\alpha \approx 0.5$, $\delta \langle E \rangle / \delta C$ is
vanishingly small indicating no significant change in the number of extreme
events even when the capacities are increased. When $\alpha > 0.7$, $\delta \langle E \rangle < 1$,
i.e, less than 1 extreme event on an average.
In this regime, capacity increase is not beneficial in alleviating the extreme events.
Since capacity addition (like building bridges, for instance) is generally an expensive proposition,
we emphasize the following important implication of this result. If the aim is to
control the mean number of extreme events on a scale-free network even while maintaining a certain
reasonable level of cost-benefit ratio, it is important to {\it apriori} estimate
if we operate in the regime in which cost-benefit ratio is favourable. The
results obtained in this work provides insights into nodal capacity addition vis-a-vis
reduction in EE. We emphasise that qualitatively similar results are obtained even when
$\Delta > 0$.

\section{conclusions}
In this work, we have studied the extreme events induced by inherent fluctuations
in the flux and taking place on scale-free networks. Given that such inherent fluctuations
are unavoidable in any finite network, we show that it is possible to tune, either
increase or decrease, the
probabilities for the occurrence of extremes in flux on a scale-free network using
nodal capacity as a tunable parameter. 
This requires that the degree distribution have a large support of at least
two orders of magnitude (see $x$-axis in Fig. \ref{fig1}). This is not true of Erdos-Renyi networks
with sharply peaked degree distributions and hence these results will not hold good for them.
As we tune the nodal capacity,
we show how various parts of the network share the burden of extreme events
and we discuss its implications. Further we study the effect of variability in
the nodal capacity.  We show that larger variability in the nodal capacity
of scale-free networks leads to more extreme events.

On the other hand, as we intuitively expect, increasing nodal capacity and hence the capacity
of the entire network leads to decrease in the mean number of extreme events in
the network. One significant result of this work is displayed in Fig. \ref{fig4}.
This shows that increasing capacity beyond certain level does not lead to proportionate
decrease in the incidences of extreme events. This has important implications for
network design efforts. Since increasing the capacity of network is an expensive proposition
in almost all the real-life situations, this work provides an important theoretical benchmark
to understand the limitations of capacity addition as a route to alleviate extreme events.
In general, real life situations involving transport on scale-free networks are generally
more complex than the random walk scenarios studied here \cite{vesp}. However, this work sets the
benchmark against which to understand extreme events arising from other types of transport
processes on networks. Further, based on evidence from earlier studies \cite{netxv}, it is possible that even
if random walk is replaced by a more intelligent routing algorithm, the results of this
work might qualitatively remain valid.

All the simulations were done on the computer cluster at IISER Pune.
ARS thanks DST for the financial support during the time this work was done.


\begin{thebibliography}{99}
\bibitem{vesp} A. Vespignani, Nature Physics {\bf 8}, 32 (2012).
\bibitem{netxv} Vimal Kishore, M. S. Santhanam and R. E. Amritkar, Phys. Rev. Lett. {\bf 106},
188701 (2011).
\bibitem{netxv1} Vimal Kishore, M. S. Santhanam and R. E. Amritkar, Phys. Rev. E. {\bf 85},
056120 (2012).

\bibitem{eevent1} A. S. Sharma, A. Bunde, V. P. Dimri, and D. N. Baker (ed.),
{\it Extreme events and natural hazards : The complexity perspective},
Geophysical Monograph Series {\bf 196}, (AGU, Washington, D. C);
S. Albeverio, V. Jentsch and Holger Kantz (Ed.) , {\it Extreme events
in nature and society}, (Springer, 2005).

\bibitem{prod} Texas Transportation Institute's 2012 Urban Mobility Report,
D. Schrank, B. Eisele and T. Lomax, (2012). (Available at mobility.tamu.edu);
K. Fisher-Vanden, E. T. Mansur and Q. Qang, National Bureau of Economic Research
Working Papers, Working paper 17741. (Available at www.nber.org/papers/w17741).

\bibitem{control} Y. Liu, J. Slotine and A. L. Barabasi, Nature {\bf 473}, 167 (2011); 
{\it ibid}, PLOS ONE {\bf 7}, e44459 (2012);
V. Nicosia {\it et. al.}, Scientific Reports {\bf 2}, 218 (2012); 
T. Nepusz and T. Vicsek, Nature Physics {\bf 8}, 568 (2012);
M. Posfai {\it et. al.}, Scientific Reports {\bf 3}, 1067 (2013).

\bibitem{rw1}  J. D. Noh and H. Reiger, Phys. Rev. Lett. {\bf 92}, 118701 (2004);
M. Rosvall and C. T. Bergstrom, PNAS {\bf 105}, 1118 (2008).
N. Zlatanov and L. Kocarev, Phys. Rev. E {\bf 80}, 041102 (2009).

\bibitem{rw2} L. F. Costa and G. Travieso, Phys. Rev. E. {\bf 75}, 016102 (2007);
V. Tejedor {\it et. al.}, Phys. Rev. E. {\bf 80}, 065104(R) (2009).

\bibitem{mss} J. F. Eichner {\it et. al.}, Phys. Rev. E {\bf 75}, 011128 (2007); M. S. Santhanam
and Holger Kantz, Phys. Rev. E {\bf 78}, 051113 (2008).

\bibitem{ym} S. Meloni et. al., Phys. Rev. Lett. {\bf 100}, 208701 (2008).

\bibitem{barrat} A. Barrat et. al., PNAS {\it 101}, 3747 (2004).

\bibitem{motter1} Dong-Hee Kim and A. Motter, New J. Phys {\bf 10}, 053022 (2008).

\bibitem{abs}  M. Abramowitz and I. A. Stegun, {\it Handbook of Mathematical Functions} (Dover, New York, 1964).

\bibitem{jam} A. Arenas, A. Diaz-Guilera and R. Guimera, Phys. Rev. Lett. {\bf 86}, 3196 (2001);
D. De Martino, L. Dall'Asta, G. Bianconi and M. Marsili,  Phys. Rev. E {\bf 79}, 
015101(R) (2009); B. Tadic, G. J. Rodgers and S. Thurner, Int. J. Bif. and Chaos {\bf 17}, 2363 (2007);
A. S. Stepanenko {\it et. al.}, EPL {\bf 100}, 36002 (2012).

\bibitem{betcen} Most of the work that uses shortest-path routing algorithms assume typical size of flux through
the nodes to be proportional to betweenness centrality. P. Holme and B. J. Kim, Phys. Rev. E {\bf 65},
066109 (2002); G. Yan {\it et. al.}, Phys. Rev. E {\bf 73}, 046108 (2006).

\bibitem{cap} W. L. Tan, F. Lam and W. C. Lau, IEEE Trans. Mob. Comput. {\bf 7}, 737 (2008).

\bibitem{cec} G. Zhang, EPL {\bf 89}, 38003 (2010).
\end{thebibliography}
\end{document}